\documentclass[iop]{emulateapj}
\usepackage{natbib}
\usepackage{color}
\usepackage{soul}              

\shorttitle{Multiple evaporating flare sources}
\shortauthors{Graham  \& Cauzzi}

\begin{document}


\title{Temporal evolution of multiple evaporating ribbon sources in a solar flare}


\author{D.R. Graham\altaffilmark{1} and G. Cauzzi\altaffilmark{1}}  
\affil{INAF-Osservatorio Astrofisico di Arcetri, I-50125 Firenze, Italy.}

\email{dgraham@arcetri.astro.it}




\begin{abstract}
We present new results from the {\em Interface Region Imaging Spectrograph} showing the dynamic evolution of chromospheric evaporation and condensation in a flare ribbon, with the highest temporal and spatial resolution to date. IRIS observed the entire impulsive phase of the X-class flare SOL2014-09-10T17:45 using a 9.4 second cadence `sit-and-stare' mode. As the ribbon brightened successively at new positions along the slit,  a unique impulsive phase evolution was observed for many tens of individual pixels in both coronal and chromospheric lines. Each activation of a new footpoint displays the same initial coronal up-flows of up to $\sim$ 300 km s$^{-1}$, and chromospheric downflows up to 40 km s$^{-1}$. Although the coronal flows can be delayed by over 1 minute with respect to those in the chromosphere, the temporal evolution of flows is strikingly similar between all pixels, and consistent with predictions from hydrodynamic flare models. Given the large sample of independent footpoints, we conclude that each flaring pixel can be considered a prototypical, `elementary' flare kernel. 
\end{abstract}


\section{Introduction}
Spectral signatures of chromospheric evaporation during a flare's impulsive phase have long been observed as strongly blue-shifted components of very hot, `coronal' lines \citep[see e.g. the review by][]{Milligan:2015aa}. However, the common occurrence of a dominant stationary component, together with the blue-shifted one, remained in contrast to the predictions of theory for single flaring loops \citep[e.g.][]{Mariska:1989aa,Li:1989aa}. A likely scenario to resolve this discrepancy invokes highly filamentary structures in which many sub-resolution magnetic loops are activated in succession. Emission would be detectable only when a number of these loops emit together, and the evaporated plasma in many of them has come to rest \citep[e.g.][]{Hori:1997aa,Doschek:2005aa}. In the 1990s and 2000s spatially resolved observations from EUV spectrometers such as SOHO/CDS \citep{1995SoPh..162..233H}, and \emph{Hinode}/EIS \citep{2007SoPh..243...19C}, started providing support to this idea. A few authors reported the occurrence of fully blue-shifted high temperature lines within flaring footpoints, corresponding to up-flows of $\sim$150--250 ${\rm km~s^{-1}}$ \citep[e.g.][]{Brosius:2013aa}. Yet, such occurrences remained sporadic in the literature.

The low-temperature counterpart of evaporation, required to conserve momentum \citep[chromospheric condensation,][]{Fisher:1987aa}, has also been observed numerous times by the CDS and EIS spectrometers, although the temperature at which downflows are observed varies greatly depending on the flare observed \citep{Milligan:2006ab,2009ApJ...699..968M,2011A&A...532A..27G}. To date, the most comprehensive studies of the spatio-temporal evolution of chromospheric condensation have been performed using ground-based observations of optically-thick chromospheric lines such as H$\alpha$ or Ca~{\sc ii} \citep[e.g.][]{Canfield:1990aa}. For example, the high resolution observations of \citet{Falchi:1997aa} proved that chromospheric condensation occurs only in very small areas ($<$1\arcsec) at the leading edge of a flaring ribbon, in agreement with the filamentary scenario outlined above.


 
The Interface Region Imaging Spectrograph \citep[IRIS,][]{2014SoPh..289.2733D} is rapidly providing new insights into this topic. With simultaneous imaging and spectroscopy in the near-UV, IRIS covers a variety of continua and emission lines formed over the chromosphere ($\sim10^4$ K), transition region and corona. Equally important, its high spatial, temporal and spectral resolution will provide for more stringent testing for models of flare heating and chromospheric dynamics. Indeed, already several IRIS papers have documented fully blue-shifted profiles of the coronal Fe~{\sc xxi} line in flares \citep{Tian:2014aa,Sadykov:2014aa,2015ApJ...799..218Y,Polito:2015aa,Tian:2015aa}. 

In this letter we use high cadence IRIS data to provide a comprehensive picture of chromospheric evaporation and condensation, tracking the entire impulsive phase evolution of tens of unique, sequentially activated, ribbon sources in an X-class flare.

\begin{figure*}
  \centering
  \includegraphics[width=18cm]{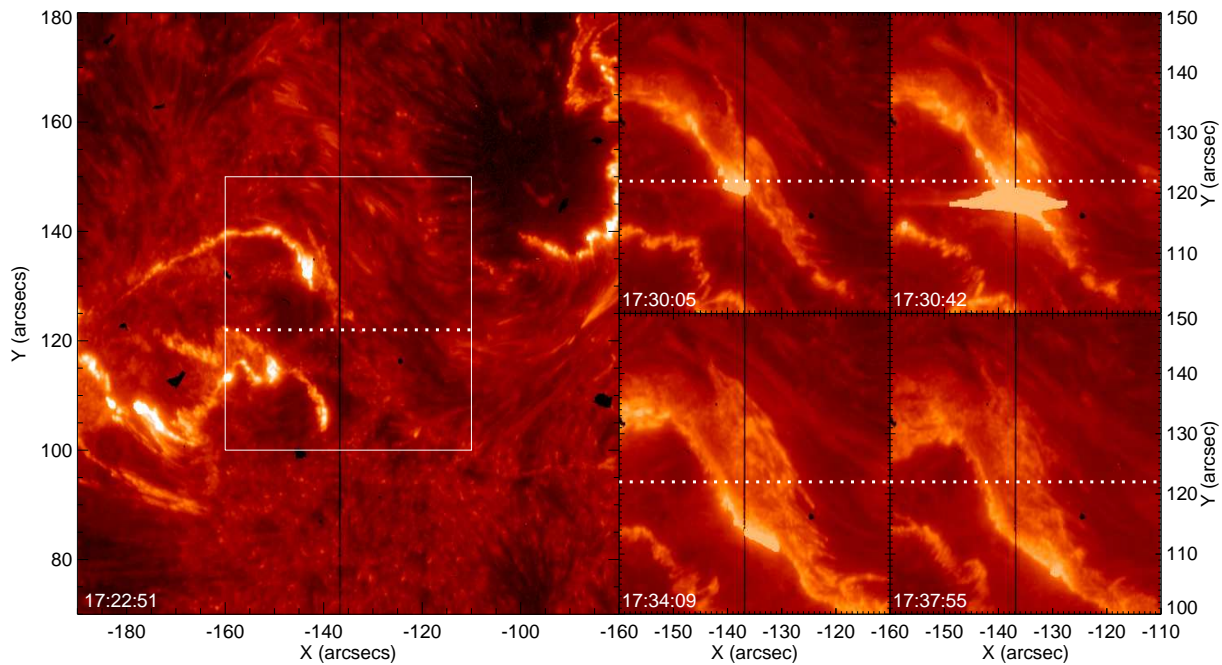}
   \caption{IRIS 1400\AA\ slit-jaw images (SJI) showing the evolution of the flaring region. The full field-of-view is seen in the larger left panel, and the four smaller panels display the ribbon evolution in the central area (white box) during the impulsive phase. Note that the 1400\AA\ SJI images saturate at times but the spectra analysed remain below this level. }
  \label{fig:sj1400}
\end{figure*}

\section{Data analysis}\label{sec:data}

The GOES X1.6 class flare SOL2014-09-10T17:45 developed in AR NOAA 12158 near disk centre (N15E02), with a complex two-ribbon structure that encompassed the main, leading polarity sunspot, and a group of several smaller spots of following polarity (see Figs. \ref{fig:sj1400} and \ref{fig:geometry}, and the studies of \cite{Li:2015ab} and \citet{Tian:2015aa}). During the impulsive phase, a rapid increase of the 20 -- 100 keV hard X-ray flux was observed by Fermi \citep{2009ApJ...702..791M} starting from $\sim$ 17:22 UT, peaking around 17:36 UT. We focus on plasma dynamics during this early impulsive phase. 

A flare watch program was run by IRIS on NOAA 12158 from 11:28 -- 17:58 UT, using sit-and-stare (SNS) mode and the standard IRIS flare line list (OBSID 3860259453). The exposure time throughout was 8 s for the FUV channel, while in the NUV it was dropped from 8 s to 2.4 s at 17:27~UT; the cadence was kept a constant 9.4 s for both channels, one of the highest ever for such a study.

The left-hand panel of Figure \ref{fig:sj1400} represents the full IRIS 1400\AA\ slit-jaw image at 17:22~UT, showing the early development of the ribbons. The vertical black line indicates the spectrograph slit. The four smaller panels on the right show the central region for different times during the impulsive phase. Comparing with the horizontal dashed line at $y=122$\arcsec\ demonstrates how the ribbon developed southwards along a portion of the slit, and the successive brightening of individual footpoint kernels over previously undisturbed regions.

\subsection{Fe XXI analysis}

A single high temperature line, Fe~{\sc xxi} 1354.1 \AA\ formed at $\sim$10~MK, is found in the IRIS spectral windows \citep[see][]{2015ApJ...799..218Y}. We prepared the data using the \emph{Solarsoft} {\sc IRIS\_GETWINDATA} procedure which returns the spectrum in photon counts with uncertainties. A spatial binning of $\pm 1$ pixels along the slit direction was used to improve the signal to noise ratio; the Fe~{\sc xxi} data shown hence have a spatial scale of 0.33\arcsec (slit width) $\times~0.5$\arcsec.

\begin{figure}
 \includegraphics[width=9cm]{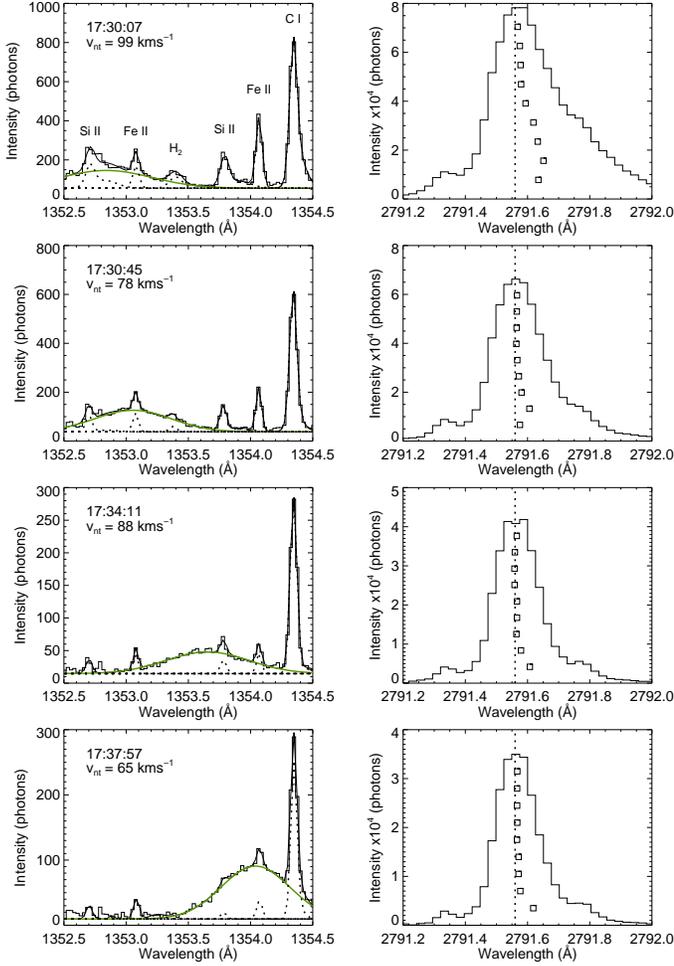}
   \caption{Left panels: fits of the Fe {\sc xxi} 1354.1\AA\ line (green) for slit position $y=122$\arcsec. The observed spectrum is shown by the stepped black line, with the total fit in black on top; cooler components are shown by a dotted line. The non-thermal FWHM, $v_{nt}$, is also shown. Right panels: Mg {\sc ii} subordinate line with bisector positions at 10\% intensity increments marked by black squares, and the rest wavelength (2791.56 \AA) by the vertical dashed line.}
  \label{fig:spec}
\end{figure}

The UV Fe~{\sc xxi} emissivity is generally low, yet the 1354.1\AA\ line has been clearly observed during flares in earlier 
observations \citep{Mason:1986aa,Cheng:1990aa}. In our X-class event we detect emission during the entire flare 
evolution, but the concomitant presence of flare-enhanced lines from cooler ions, including Si~{\sc ii}, Fe~{\sc ii}, 
C~{\sc i} requires special care in order to extract the Fe~{\sc xxi} signal. We make Gaussian fits to the strong lines present in the 1352.4 -- 1355.9 \AA\ spectral window (labeled in Fig. \ref{fig:spec}), allowing for the presence of a second, red-shifted component of the cooler lines when necessary. The fit was further constrained by locking the centroids and widths of the Si~{\sc ii} and Fe~{\sc ii} lines to the same species present in the adjacent 1347.8 - 1350.9 \AA\ spectral window.


In Figure \ref{fig:spec} the fit to the Fe~{\sc xxi} line is shown for slit position $y=122$\arcsec\ and the 4 time steps of the right-hand panels of Figure \ref{fig:sj1400}.  In the earliest frame the Fe~{\sc xxi} line is shifted almost to the blue edge of the spectral window, with an up-flow velocity of 280~${\rm km~s^{-1}}$; its deceleration is easily tracked in time moving towards the rest wavelength at 1354.1\AA.

A similar fit was performed over all of the flaring slit pixels. In most of them, the earliest instance of the line is extremely broad and of comparable intensity to the background chromospheric lines; to avoid mis-identification, we have tracked the Fe~{\sc xxi} enhancement back in time by examining the series of spectra by eye, determining when the flow speed stops increasing, or the line is otherwise undetectable. Fits prior to this cut-off are not included in the analysis. Nowhere do we find evidence for a separate `stationary component' of the Fe~{\sc xxi} line: the chromospheric lines are well identified and there is no significant residual between the fit and data to suggest that one exists.


\subsection{Mg II analysis}

The Mg~{\sc ii} resonance h\&k lines, as well as their subordinate triplet, are among the best chromospheric diagnostics within the IRIS spectral range, even for flaring conditions \citep{Leenaarts:2013aa,Pereira:2015aa}. As the h\&k lines saturated in several of the brightest flaring pixels within our flare, we identified the subordinate 2791.6 \AA\ as the best candidate for studying chromospheric condensation: the line is several \AA\ removed from the Mg~{\sc ii} k, hence less influenced by its variations during flares than the self-blended component at 2798.7 \AA.

The data was prepared using the same method as Fe~{\sc xxi}, however, given the stronger signal we maintain the original pixel scale of $0.33\arcsec \times 0.166$\arcsec. The rest wavelength was determined by averaging the line core position in a non-flaring region over the multi-hour observation, and accounting for the spacecraft orbit variation; we estimate the velocity zero point to $\pm0.7~\rm{km~s} ^{-1}$.

As proven by ground-based observations, useful indications about the amplitude and evolution of the chromospheric condensation can be obtained from optically thick lines by measuring their bisector, i.e. the locus of mid-points measured at different intensity levels within the line \citep[see][]{Ding:1995aa}. In Fig. \ref{fig:spec}, right panels, we show the Mg~{\sc ii} 2791.6 \AA\ line profiles as observed in the same pixel and times as the corresponding Fe~{\sc xxi} panels. The shape and characteristics of the Mg~{\sc ii} lines are consistent with older results \citep{Canfield:1990aa,Ding:1995aa}: almost all the flaring pixels initially display a red asymmetry of the line, which disappears rapidly; the peak of emission is only slightly red-shifted; and the shift of the bisector with respect to the rest wavelength appears to increase from the centre towards the wing, perhaps signifying a gradient in the condensation velocity \citep{Cauzzi:1996aa}. To derive the condensation velocity, we interpret the bisector position at the 30\% intensity level in terms of Doppler shifts. While this might underestimate the actual condensation velocity \citep[e.g.][]{Canfield:1990aa}, it is the best compromise between assessing the flows whilst avoiding contamination from possible blends in the far wings, although these are mostly of photospheric origin (Pereira, private comm.). We confirm that similar shifts are derived when using the other triplet lines, albeit with more scatter.


\section{Plasma dynamics}
\begin{figure}
  \centering
  \includegraphics[width=8.5cm]{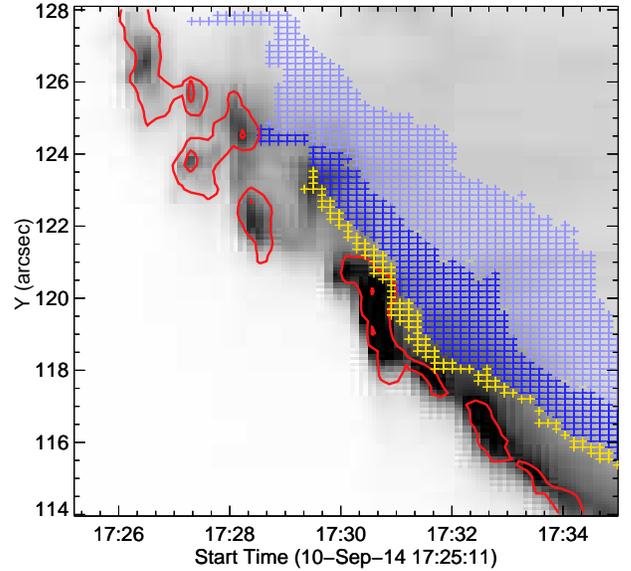}
   \caption{Mg {\sc ii} intensity space-time map (inverted B/W colour table) with overlays of Mg {\sc ii} down-flows at the 30\% bisector level at 15 and 30~$\rm{km s} ^{-1}$ (red contours).  The Fe {\sc xxi} up-flow velocities above 270, 200 and 100 $\rm{km s} ^{-1}$ are indicated with yellow, dark blue, light blue crosses, respectively.} 
  \label{fig:map}
\end{figure}

Figure \ref{fig:map} shows a space-time map for a section of the slit (81 original pixels) where the bright ribbon was found to expand southward at a steady rate of $\sim$ 0.03\arcsec\ s$^{-1}$. These pixels have the cleanest temporal evolution as the ribbon does not dwell over any one of them for longer than 1-2 time steps. The spatial offset between the Mg~{\sc ii} and Fe~{\sc xxi} detectors was corrected by aligning the spectrograph's fiducial marks.

The Mg~{\sc ii} intensity shows a sudden enhancement due to the flare, and the development of the ribbon clearly appears as a diagonal strip across the diagram. The chromospheric downflows are co-spatial, and co-temporal with each of the new intensity enhancements, as shown by the red contours at the 15 and 30~${\rm km~s^{-1}}$ level. These values of condensation flows are consistent with results from many earlier ground and space-based studies but Figure \ref{fig:map} offers unprecedented detail on their spatial and temporal evolution.

The corresponding evaporation velocities, as derived from the centroid of the Fe~{\sc xxi} fits, are overlaid as coloured crosses on the same figure. For about 70\% of the pixels we find flows reaching nearly 300~${\rm km~s^{-1}}$ (yellow crosses identify flows above 270~${\rm km~s^{-1}}$); these encompass the fastest, early Fe~{\sc xxi} emission, pertaining to a very thin spatio-temporal strip approaching our 0.5\arcsec\ resolution, and lasting only 1-2 temporal steps. Such values of evaporation flows are among the strongest ever reported \citep[300-400 km s$^{-1}$, e.g.][]{1982SoPh...78..107A}, and the largest documented yet from IRIS. Within the same pixels, the flows decay to 200~${\rm km~s^{-1}}$ (dark blue region) and below rather uniformly in time. For pixels northward of 124\arcsec, we detect only slower flows, up to $\sim 150~ {\rm km~s^{-1}}$.



\begin{figure}
  \centering
  \includegraphics[width=8.5cm]{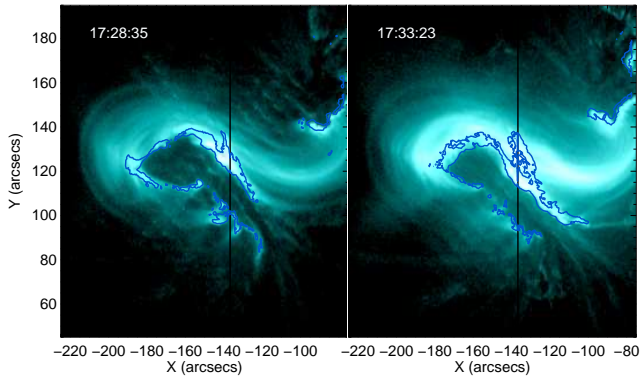}
   \caption{SDO/AIA 131 \AA\ images at two times during the impulsive phase, highlighting the geometry of the
   hot flare loops. The contours show the IRIS SJI 1400\AA\ 800DN level.}
  \label{fig:geometry}
\end{figure}

From Figure \ref{fig:map} we see that all flaring pixels display clear signatures of both evaporation and condensation; yet the onset of opposite flows is co-temporal (within 1-2 time steps) for only a few; for most pixels the initial coronal upflow lags behind the condensation by a minute or more. 
The delay could be due to a number of reasons: the Fe~{\sc xxi} line could be so blue-shifted to fall outside of the detector's edge; or the initial emission be too weak to be detected at its earliest inception; or the same-pixel Mg~{\sc ii} and Fe~{\sc xxi} signatures could derive from distinct flare loops, depending on their orientation with respect to the line-of-sight. Yet, this appears unlikely because of the flare's geometry and evolution: SDO/AIA 131 \AA\ images (mostly showing Fe~{\sc xxi} emission during flares \citep{Simoes:2015aa}) clearly show that the hot flare loops display a dominant E-W orientation, with little line-of-sight superposition of the loops and ribbons (Fig. \ref{fig:geometry}).
While plausible that a previously activated hot loop could `cloud' a new ribbon kernel developing southward of it, this should lead to Fe~{\sc xxi} flows appearing at that spatial pixel {\it before} the chromospheric flows, contrary to what was observed.


We also note that similar delays ($\sim$ 60--75 s) between chromospheric and coronal flows have been reported for other flares \citep[][Brosius 2015, private comm.]{2015ApJ...799..218Y}.


\section{Temporal Evolution of Flows}

\begin{figure*}
  \centering
  \includegraphics[width=12cm]{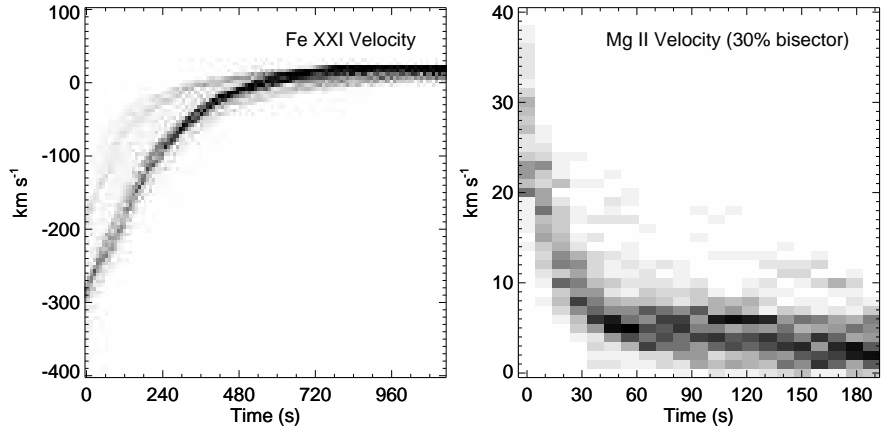}
   \caption{Superposed epoch analysis of Fe {\sc xxi} and Mg {\sc ii} flows for every slit pixel in Figure \ref{fig:map} (negative velocities showing rising material). The greyscale darkens with increasing occurrence within a given velocity interval (see text for full detail). }
  \label{fig:curves}
\end{figure*}

In Figure \ref{fig:curves} we show the full evolution of both Fe and Mg velocities for every slit position of Figure \ref{fig:map}. To highlight commonalities in the evolution we perform a superposed epoch analysis by shifting the curves to a common time origin. As the spatio-temporal coincidence of flows between both lines is not certain (Fig. \ref{fig:map}), we use a different time origin for each: the Fe {\sc xxi} curves (left panel) are shifted so that the earliest detectable fit occurs at ${t=0}$, while in Mg~{\sc ii} (right panel) the velocity curves are aligned to their peak time. For each time step, the number of spatial pixels with velocity between $\delta v = 5~{\rm km~s^{-1}}$ (Fe~{\sc xxi}) and $\delta v = 1~{\rm km~s^{-1}}$ (Mg~{\sc ii}) is represented by the greyscale, with a maximum (black) of 20 occurrences. 

In Fe~{\sc xxi} all of the pixels experience an instantaneous appearance of the fastest upflows, and although the centroids for the Fe~{\sc xxi} fit are allowed freedom over the entire IRIS window (a $\sim380~{\rm km~s^{-1}}$ interval), there is very little random variation in velocity between time steps (the averaged uncertainty drops from $\pm15$ to $\pm2~{\rm km~s^{-1}}$ within 50 s). The majority of pixels fall consistently on a very narrow curve, with an initial velocity around 300 km s $^{-1}$ and a decaying upflow lasting over 6 minutes. The smaller grouping of pixels to the north of Figure \ref{fig:map}, have instead slower initial flows and plateau earlier, but their decay curve is comparable in shape. 

The evolution of evaporation upflows in spatially resolved sources has been discussed only in a handful of papers\footnote{We note that similar velocity evolution curves have been observed for the case of two X-class flares by \citet{2011SoPh..273...69H} using disk-integrated EVE data.}. In the case of a single flaring footpoint \citet{Brosius:2013aa,Polito:2015aa,Tian:2015aa} all found velocity curves similar to those displayed in Fig. \ref{fig:curves}, but with decay times ranging between $\sim$3 - 8 minutes, possibly due to differences in the respective data's temporal (9.4 - 75 s) and spatial sampling. Given the large number of individual flaring kernels represented, the high cadence of the measurements, and the homogeneity of the results, we think that Figure \ref{fig:curves} represents the best picture to date of hot evaporating flaring plasma.


In Mg~{\sc ii}, strong condensation downflows are apparent, reaching nearly 40~${\rm km~s^{-1}}$. Again, most pixels fall on a very consistent curve, which decays rapidly and approaches near zero velocity in only $\sim 50-60$ s. This is in very good agreement with predictions from theoretical models of chromospheric condensation; for example, for typical pre-flare chromospheric conditions \citet{Fisher:1989aa} indicates 1 minute as the necessary time for the condensation to stop, essentially independent from the flare energy input. Most of the previous condensation studies found a much longer decay time, around 2-3 minutes \citep{Ichimoto:1984aa,Ding:1995aa}. It is likely that their results depended on a superposition of unresolved events, occurring successively within their spatial resolution elements (typical pixel size of $\sim$ 2--3\arcsec).

\section{Discussion and Conclusions}\label{sec:discussion}

We have presented a unique IRIS dataset, covering the development of a large portion of a ribbon during the impulsive phase of an X-class flare. The favourable positioning of the spectrograph slit allowed us to observe several tens of resolved flaring areas, and to follow their complete temporal evolution at high cadence. In this letter we concentrated on their dynamical properties; our main conclusions are as follows:

1) All of the flaring pixels (at $\sim 0.5$\arcsec\ resolution) display sudden and strong Fe~{\sc xxi} upflows at the beginning of the flare, and we find that the coronal line is {\em always} entirely blue-shifted. This is consistent with other recent IRIS results and seems to imply that IRIS observations fully resolve single flaring kernels. Still, upflows persist in any given position for a remarkable length of time, and should be readily visible even at the coarser resolution of CDS or EIS (cf. Fig. \ref{fig:map}), contrary the commonly reported occurrence of a dominant stationary component in such observations. A possible explanation might reside in the very simple linear progression of the newly activated kernels analysed. However, we cannot state if these observed characteristics are representative of the entire flare or of other events.



2) All of the flaring pixels (at $\sim 0.3$\arcsec\ resolution) also display sudden and strong Mg~{\sc ii} condensation downflows, with values in agreement with earlier results from both visible and EUV observations. The chromospheric condensation in each flaring kernel stops in $\sim 50-60$ s, at least a factor of two faster than any previously reported value, but consistent with predictions of 1-D hydrodynamical simulations of flares affecting `undisturbed' chromosphere.

3) Surprisingly, only a few pixels show a simultaneous onset of coronal and chromospheric flows, while for most of them the initial coronal evaporation lags behind the chromospheric condensation by an average of 68 seconds. This appears contrary to the standard explosive scenario. From an analysis of the coronal loop geometry, line-of-sight superposition effects do not seem sufficient to explain this delay. As the same trend has been reported in other recent IRIS studies we speculate that the delay could simply be caused by the Fe~{\sc xxi} emission being too weak to be detected at its earliest inception.

4) The Fe~{\sc xxi} spectra are extremely broadened compared to the ion's thermal FWHM of 92~${\rm km~s^{-1}}$ (see Figure \ref{fig:spec}). \citet{Polito:2015aa} find similar excess broadening and discuss its potential origins, including a plasma temperature beyond the equilibrium formation temperature, and unresolved plasma motions; the later seeming probable when considering the initial rapid change in velocity (Figure \ref{fig:curves}) for individual pixels.

5) Figure \ref{fig:curves} represents the clearest picture to date of the temporal evolution of both chromospheric evaporation and condensation. The evolution of plasma dynamics is so strikingly similar for most of the pixels, it suggests that the characteristics of the energy release are either remarkably uniform, each time occurring in a pristine environment, or have little influence over the subsequent plasma evolution.

We conclude by remarking that the large number of independent flaring pixels observed, and the complete temporal coverage of their dynamical evolution at high cadence allow us to derive common characteristics of what we can define as `prototypical' flares, with a spatial extension limited by the actual resolution of our data, i.e. $\le$ 0.5\arcsec. In principle our results can be immediately compared with the output of numerical simulations of single flaring loops. For example, with respect to the third point above, one could attempt to derive values of the actual coronal emission during the early phases of flare chromospheric heating, as predicted within either collisional thick-target \citep{2005ApJ...630..573A} or conductive \citep{Longcope:2014aa} models.

\acknowledgments
This research has received funding from the European Community's Seventh Framework Programme (FP7/2007-2013) under grant agreement no. 606862 (F-CHROMA). IRIS is a NASA small explorer mission developed and operated by LMSAL with mission operations executed at NASA Ames Research center and major contributions to downlink communications funded by the Norwegian Space Center (NSC, Norway) through an ESA PRODEX contract. We thank  L. Fletcher, H. Hudson, R. Falciani, and the two referees for their helpful comments and discussion.

\bibliographystyle{apj}

\end{document}